# A Nuclear Periodic Table

K. Hagino and Y. Maeno

Department of Physics, Kyoto University, Kyoto 606-8502, Japan

E-mail:

K.Hagino

hagino.kouichi.5m@kyoto-u.ac.jp

Y. Maeno

maeno.yoshiteru.2e@kyoto-u.ac.jp

ORCID

K. Hagino        0000-0002-2250-1063

Y. Maeno         0000-0002-3467-9416

**Abstract**

There have been many empirical evidences which show that the single-particle picture holds to a good approximation in atomic nuclei. In this picture, protons and neutrons move independently inside a mean-field potential generated by an interaction among the nucleons. This leads to the concept of nuclear shells, similar to the electronic shells in atoms. In particular, the magic numbers due to closures of the nucleonic shells, corresponding to noble gases in elements, have been known to play an important role in nuclear physics. Here we propose a periodic table for atomic nuclei, in which the elements are arranged according to the known nucleonic shells. The nuclear periodic table clearly indicates that nuclei in the vicinity of the magic numbers can be understood in terms of a shell closure with one or two additional nucleons or nucleon holes, while nuclei far from the magic numbers are characterized by nuclear deformation.

**Introduction**

Atomic nuclei are located at the center of atoms and carry almost all the fraction of the mass of atoms. They consist of a small number of protons and neutrons, collectively called nucleons. It has been known that many properties of atomic nuclei can be understood in a simple mean-field approximation, in which nucleons move independently from each other in a common potential, that is, a mean-field potential generated as a consequence of the interaction among nucleons (see e.g., Bohr and Mottelson 1969; Ring and Schuck 1980). In this picture, nucleons occupy single-particle orbits, like electrons in atoms, which naturally leads to the concept of shell structure and shell closures. Such shell structure in atomic nuclei have been evidenced by many phenomena, such as an increased binding energy, discontinuities of the nuclear radius as well as neutron and proton separation energies (corresponding to the ionization energy in atoms), and an increased excitation energy of the first excited state, all of which occur at the shell closures. For stable nuclei, it has been known that the shell closures occur when the number of neutron or proton takes the so called magic numbers given by 2, 8, 20, 28, 50, 82, and 126. These magic numbers have been explained using a finite depth potential with a strong spin-orbit interaction, as is shown in Fig. 1 (Mayer 1949; Haxel *et al*. 1949).

Since magic nuclei, that is, nuclei with a shell closure of neutron or proton, correspond to noble gases in the elements, one may expect a similar periodic table for nuclei to the one for the elements. The aim of this paper is to present such nuclear periodic table, based on the shell structure of atomic nuclei. Of course, atomic nuclei consist of two different kinds of particles, that is, proton and neutron, and thus their properties cannot be determined solely by the number of protons $Z$ (or of neutrons $N$). As a matter of fact, a two-dimensional map on the $N$-$Z$ plane, known as a nuclear chart, is widely known for atomic nuclei (Koura *et al*. 2018; Magil *et a*l. 2018), which can be used to learn e.g., the abundance fraction of each stable nucleus and a lifetime of each unstable nucleus. Such map would be more useful than a one-dimensional ($Z$ or $N$) periodic table in order to classify detailed properties of atomic nuclei. Nevertheless, it might still be useful and pedagogical to construct a nuclear periodic table e.g., in order to visualize the difference in magic numbers between nuclear systems and electronic systems. In this regard, we would like to point out that the occurrences of the magic numbers are not explicitly presented in a standard nuclear chart, which shows only the abundance ratio or the decay half-life for each nucleus.

In this paper, we construct a nuclear periodic table based on the shell structure of protons. Even though one could construct a similar periodic table for neutrons, it is much easier to have a nuclear periodic table based on protons, as each proton magic number can be characterized by the name of a particular element. We notice that a similar nuclear periodic table has recently been proposed (Więckowski 2019). However, in the nuclear periodic table of Więckowski, the magic numbers are placed in different columns; thus, the correspondence between the atomic and the nuclear periodic tables is not very clear. In the nuclear periodic table proposed in this paper, we place the magic numbers in the same column, as in the noble gases in the periodic table of elements, and it is more straightforward to see differences and similarities between atomic and nucleonic systems.

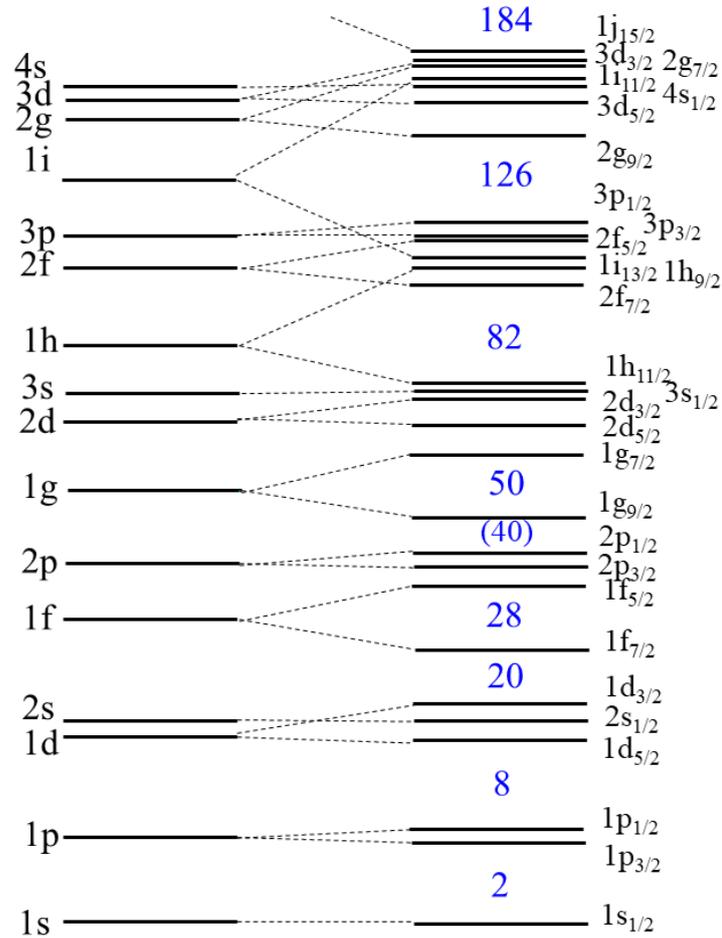

**Fig.1**

Typical single-particle levels of nucleons in a spherical mean-field potential. These are labeled by the quantum numbers $n$, $l$, and $j$, where $n$ is the radial quantum number, that is, the number of node (plus 1) in the radial wave function, $l$ is the orbital angular momentum, and $j$ is the total angular momentum due to the addition of the orbital and the spin angular momenta. The notations $s, p, d, f, g, h$…. correspond to $l$ = 0, 1, 2, 3, 4, 5…, respectively. Note that the radial quantum number $n$ is different from the principle quantum numbers, $n_p=n+l$, which are often used to label electronic single-particles levels. The degeneracy of each single-particle level is given by $2j+1$. Because of a strong spin-orbit interaction, a single particle level with an orbital angular momentum $l$ splits into two levels with the total angular momentum of $j = l \pm 1/2$, among which the level with $j = l + 1/2$ is lowered in energy than the level with $j = l - 1/2$. The ordering of the levels within each shell may be altered depending on the number of protons and neutrons. The magic numbers due to the shell closures are also indicated.

**Nuclear periodic table**

In order to construct a nuclear periodic table, we first arrange the elements with the proton magic numbers in the same column. Those are: He ($Z=2$), O ($Z=8$), Ca ($Z=20$), Ni ($Z=28$), Sn ($Z=50$), and Pb ($Z=82$). Zr ($Z=40$) often shows behaviors similar to the magic nuclei due to the sub-shell closure at $Z=40$ (see Fig. 1; Garcia-Ramos and Heyde 2019), and we also include it in the same column. Even though the magic numbers may change in neutron-rich nuclei when the number of neutrons is much larger than the number of protons (Steppenbeck 2013; Otsuka 2020), in this paper we consider only those nuclei which are close to the beta-stability line and do not consider such effect. Though the heaviest element discovered so far is Oganessson ($Z=118$), the proton magic number after $Z=82$ is currently unknown. Theoretical calculations based on the so called macroscopic-microscopic approach carried out in the 1960s have predicted that the proton shell closure deviates from $Z=126$ due to the Coulomb interaction among protons, and appears at $Z=114$ together with the neutron shell closure at $N=184$ (Sobiczewski *et al*. 1966; Nilsson *et al*. 1968). Here, the proton shell closure at $Z=114$ is obtained by filling the $2f_{7/2}$, $1h_{9/2}$, and $1i_{13/2}$ orbitals above the $Z=82$ magic number (see Fig. 1). The region around this nucleus, $^{298}_{114}Fl_{184}$, has been called the island of stability, providing an important motivation to explore superheavy elements. More recent calculations have predicted different proton shell closures ($Z=114$, 120, or 126) depending on a theoretical model employed, even though the neutron shell closure at $N=184$ is more robust (Bender *et al*. 1999). In this paper, we choose the traditional proton magic number, $Z=114$, for superheavy elements and arrange Fl underneath Pb in the nuclear periodic table.

After we set up the column for the magic and semi-magic nuclei, we next arrange other nuclei according to the nuclear shell structure shown in Fig. 1. The ordering of each single-particle level within shells depends on the number of neutrons. Moreover, for open shell nuclei, those single-particle levels are occupied only partially due to the pairing correlation. In mid-shell nuclei, nuclei may even be deformed, yielding a deformed mean-field potential. In that situation, the single-particle levels shown in Fig. 1, which are based on a spherical mean-field potential, lose their clear physical meaning. We therefore consider a group of single-particle levels within each shell, instead of treating each single-particle level individually.

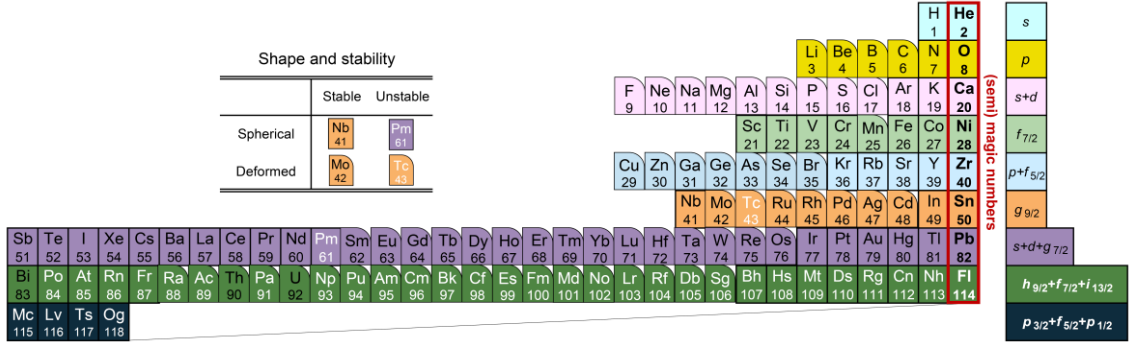

**Fig.2**

A nuclear periodic table based on the proton magic numbers. The rightmost column shows the elements with the proton magic and semi-magic numbers. The other elements are arranged according to the nuclear shell structure shown in Fig. 1, for which the single-particle levels for the valence protons are denoted with different colors. In the legend for the single-particle levels, those without $j$ include both of the spin orbit partners, e.g. $f$ for $f_{5/2}+f_{7/2}$. The elements shown in round-corner boxes are those whose nucleus is deformed in the ground state (see Möller e*t al*. (Möller e*t al*. 2016) for the actual values for the deformation parameter). Elements with black symbols have stable nuclei, while those with white symbols represent those with all the isotopes unstable. Even though Bi and Th are unstable, we do not include them in the unstable elements since their decay half-lives are of the order of the age of the universe or longer. Likewise, we do not include U in the unstable elements, since the half-life is similar to the age of the earth.

Figure 2 shows a nuclear periodic table so constructed. The elements shown in round-corner boxes are the ones in which the nucleus is statically deformed in the ground state. Here, we regard that a nucleus is deformed when the absolute value of the quadrupole deformation parameter, $\beta_2$, is larger than 0.15. The deformation parameter $\beta_2$ is related to the angle dependent radius of a nucleus given by

$$R(\theta) = R_0 \left(1+\beta_2\, Y_{20}(\theta)\right),$$

where $R_0$ is the radius of the sphere and $Y_{20}$ is the spherical harmonics. Here, we have assumed that a nucleus has axial symmetric shape and took the symmetric axis to be the $z$-axis ($\theta = 0$). For each element, we choose the most abundant nucleus and estimate the deformation parameter using the theoretical calculations by Möller *et al*. (Möller *et al*. 2016). We note that the resultant periodic table will be almost the same even if we choose the deepest bound isotopes. For the elements lighter than N, we regard the elements Li,

Be, B, and C as deformed due to the well-known alpha-particle structure of atomic nuclei. The elements with the white symbols are unstable elements, that is, elements where all the isotopes are unstable.

In the figure, one can immediately see that the elements in the vicinity of the shell closures are all spherical, while the deformation is developed in the mid-shell regions. The former elements can be interpreted in terms of one or two protons holes outside the shell closures, and it may be meaningful to arrange them in the same columns.

Because of particle-hole symmetry, it may be more appropriate to put the elements symmetrically around the shell closures. This is done in Fig. 3, in which the magic and the semi-magic nuclei are put in the center of the periodic table. Its paper model, inspired by a similar spiral model of the atomic periodic table (Maeno 2002), is shown in Fig. 4. One can see that many elements with one additional proton to the shell closures are spherical in the ground state, even though Li and Cu show deformation, the latter of which is caused by a correlation between the neutrons in Cu and the valence proton. For the nuclei with two valence protons outside the closed shells, the nuclei tend to be deformed due either to the alpha-particle structure (Be and Ne) or to the fact that the neutron is in the mid-shell (Zn and Mo).

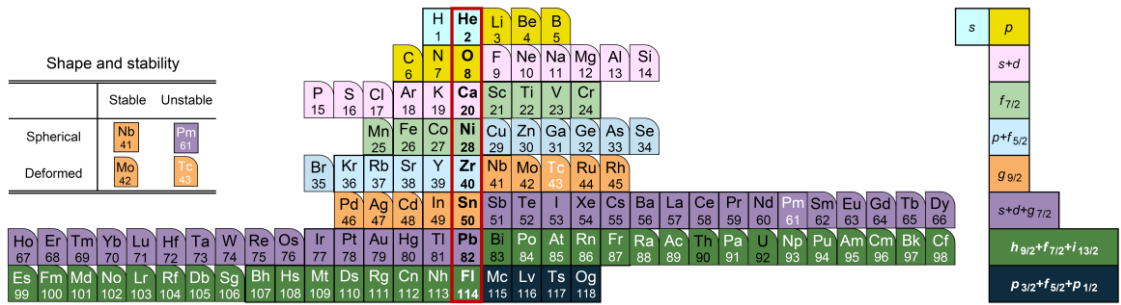

**Fig.3**
A nuclear periodic table similar to Fig. 2, but a version in which the elements are arranged symmetrically around the shell closures.

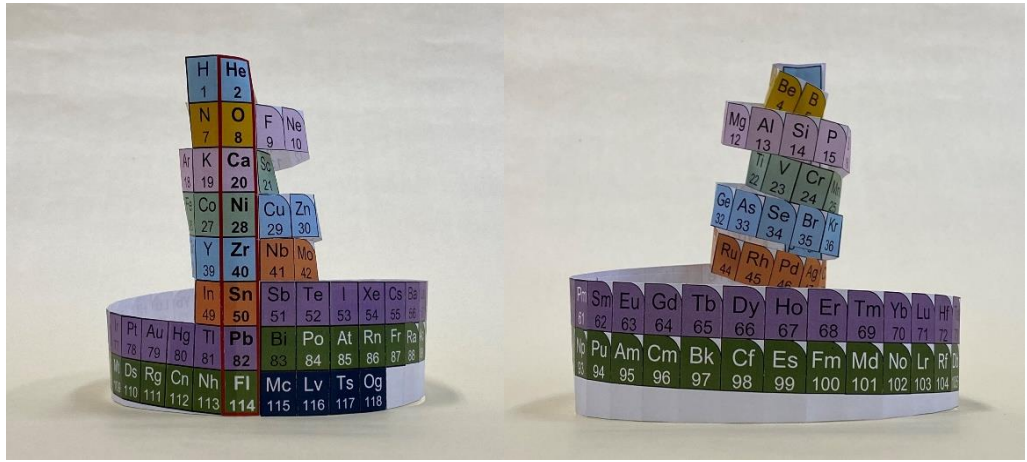

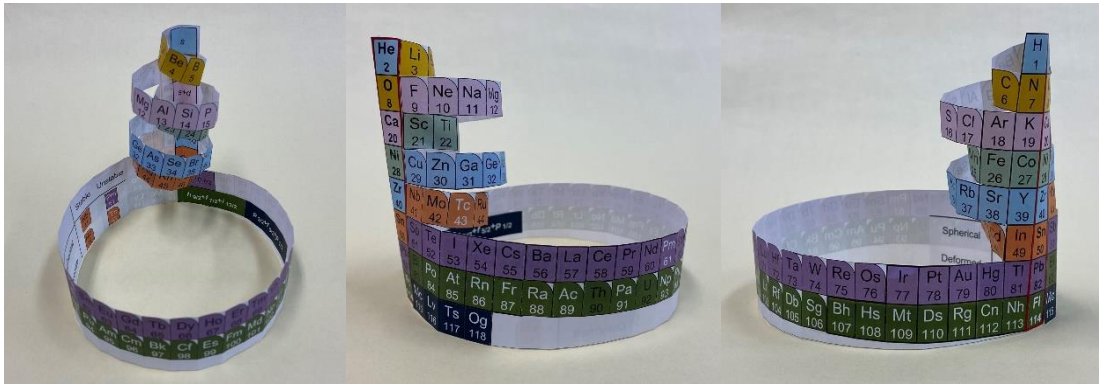

**Fig.4**

A paper model of the nuclear periodic table shown in Fig. 2. Corresponding to a similar model for the atomic periodic table known as *"Elementouch"* (Maeno 2002), we propose to call this model *"Nucletouch"*.

**Summary**

We have constructed a nuclear periodic table, based on the shell structure of protons in atomic nuclei. To this end, we have considered the nuclei close to the stability line, and arranged the elements according to the known proton magic numbers. The nuclei in the vicinity of the shell closures can be interpreted as one or two protons or proton holes outside the shell closures, and thus may have similar properties to one another. The nuclear periodic table clearly shows that those nuclei are spherical in the ground state, while nuclei tend to be deformed when the distance from the shell closures is large. The nuclear periodic table may also be useful in order to visualize the difference in magic numbers between nuclear and electronic systems.

It has been known that some of the nuclear magic numbers are changed in neutron-rich nuclei. It might be amusing to construct an extended version of nuclear periodic table by taking into account such shell evolution in exotic nuclei.

**Acknowledgements:**
We thank H. En'yo for useful discussions, and K. Takamiya and T. Kodama for help. This was partially supported by Japan Society for the Promotion of Science (JSPS) KAKENHI No. JP17H06136 and by the JSPS Core-to-Core Program.